\documentclass[nofootinbib]{revtex4}
\usepackage{graphicx}
\usepackage{latexsym}
\def\be{\begin{equation}}
\def\ee{\end{equation}}
\def\bea{\begin{eqnarray}}
\def\eea{\end{eqnarray}}
\def\bi#1{\hbox{\boldmath{$#1$}}}
\newcommand{\wjma}[6]{\left(
                           \begin{array}{ccc}
         #1 & #2  & #3  \\
         #4 & #5  & #6
                           \end{array}
                   \right)}

\begin{document}

\title{Cosmological CPT violating effect on CMB polarization}

\author{Mingzhe Li$^{1,2,3}$
and Xinmin Zhang$^{4}$ }

\affiliation{${}^1$ Fakult\"{a}t f\"{u}r Physik, Universit\"{a}t
Bielefeld, D-33615 Bielefeld, Germany} \affiliation{${}^2$
Department of Physics, Nanjing University, Nanjing 210093, People's Republic of 
China} \affiliation{${}^3$ Joint Center for Particle, Nuclear
Physics and Cosmology, Nanjing University - Purple Mountain
Observatory, Nanjing 210093, Peoples Republic of China} \affiliation{${}^4$
Institute of High Energy Physics, Chinese Academy of Sciences, P.O.
Box 918-4, Beijing 100049, People's Republic of China}


\begin{abstract}
A dark energy scalar (or a function of the Ricci scalar) coupled with
the derivative to the matter fields will
violate the $CPT$ symmetry during the expansion of the Universe.
This type of cosmological $CPT$ violation helps to generate the baryon
number asymmetry and  gives rise to the rotation of  the photon
polarization which can be measured in the astrophysical and
cosmological observations, especially the experiments
 of the cosmic microwave background radiation.  In this paper, we
 derive the rotation angle in a fully general relativistic way and present the rotation formulas
 used for the cosmic microwave background data analysis. Our formulas include the corrections from the spatial fluctuations of the scalar field.
  We also estimate the magnitude of these corrections in a class of dynamical dark energy models for quintessential baryo/leptogenesis.
\end{abstract}

\maketitle

\hskip 1.6cm PACS number(s): 98.80.Es, 98.80.Cq, 98.70.Vc \vskip 0.4cm

\section{Introduction}

In the standard model of particle physics, $CPT$ is a fundamental symmetry.
Probing its violation is an important way to search for new physics beyond the standard model. Up
to now, $CPT$ symmetry has passed a number of high precision experimental tests in the ground-based laboratory and no
definite signal of its violation has been observed \cite{Lehnert:2006mn}. So, the present $CPT$ violating effects,
if they exist, should be very small to be
amenable to the experimental limits.

The $CPT$ symmetry could be violated dynamically in the expanding Universe.
 To show it, consider a scalar boson
$\phi$ which effectively couples to a fermion current $J^\mu$, with
the Lagrangian given by \be\label{baryon} \mathcal{L}_{\rm
int}=\frac{c}{M}\nabla_{\mu}{\phi}J^{\mu}~, \ee where $c$ is a
dimensionless constant and $M$ is the cut-off scale. The interaction
in (\ref{baryon}) is $CPT$ conserved; however, during the expansion of the Universe, the
background value $\dot \phi$ does not vanish and $CPT$ is broken
spontaneously. This type of $CPT$ violation occurs naturally in
theories of dynamical dark energy and has interesting implications
in particle physics and cosmology. In models of quintessential
baryo/leptogenesis
\cite{quin_baryogenesis,k-baryogenesis,DeFelice:2002ir}, the scalar
field $\phi$ in (\ref{baryon}) is the dark energy scalar
(quintessence \cite{Wetterich:1987fm,Ratra:1987rm,Caldwell:1997ii},
$k$-essence \cite{ArmendarizPicon:2000dh}, phantom \cite{Caldwell:1999ew}, quintom \cite{Feng:2004ad,Li:2005fm} etc.). In the early
Universe the field $\phi$ with the interaction in (\ref{baryon})
generates the baryon number asymmetry and at late times it drives
the accelerating expansion of the Universe. One of the features of these
models is a unified description of the present accelerating
expansion and the generation of the matter and antimatter asymmetry
of our Universe. Furthermore, differing from the original proposal
for spontaneous baryogenesis by Cohen and Kaplan
\cite{spon_baryogenesis}, since the dark energy scalar has been
existing up to the present epoch, the corresponding $CPT$ violation could
be tested in laboratory experiments and cosmology. In Refs.
\cite{quin_baryogenesis,k-baryogenesis}, we have pointed out that, to
produce the enough baryon number asymmetry, the dark energy should
be significant in the radiation-dominated epoch. This is the case if
the dark energy has the tracking behavior, i.e., its density decays
almost at the same rate with that of radiation as the Universe
expanding. Along this line, the gravitational baryo/leptogenesis
\cite{steinhardt,lihong} has been proposed in which a function of
curvature scalar $R$ replaces the $\phi$ field in (\ref{baryon}). 
There are other motivations in the literature, e.g., Refs. \cite{Bertolami:1996cq,Kostelecky:2002ca, Bertolami:2003qs, 
Alberghi:2003ws, Mukhopadhyay:2007vca, Sinha:2007uh}.

The current $J^{\mu}$ in Eq. (\ref{baryon}) is not necessary to
be the baryon current for baryogenesis. It could be other currents
which are not orthogonal to $J^{\mu}_B$ or $J^{\mu}_{B-L}$. In
\cite{Li:2006ss}, we have proposed, for example, that $J^{\mu}$ is
the left-handed part of the $B-L$ current $J^{\mu}_{(B-L)_L}$.
Besides the generation of baryon number asymmetry, this kind of
coupling will bring a new effect to the photon sector. This is because
$J^{\mu}_{(B-L)_L}$ is anomalous under the electromagnetic
interaction \be \nabla_{\mu}J^{\mu}_{(B-L)_L}\sim
-\frac{\alpha_{em}}{3\pi}F_{\mu\nu}\widetilde{F}^{\mu\nu}~. \ee
Hence the interaction in Eq. (\ref{baryon}) would induce the
following effective coupling through the anomaly equation:
\be\label{int} \mathcal{L}_{\rm int}=-\frac{2c\alpha_{em}}{3\pi
M}\nabla_{\mu}{\phi}A_{\nu}\widetilde{F}^{\mu\nu}\equiv p_{\mu}
A_{\nu}\widetilde{F}^{\mu\nu}~, \ee where $A_{\nu}$ is the
electromagnetic vector potential,
$F_{\mu\nu}=\nabla_{\mu}A_{\nu}-\nabla_{\nu}A_{\mu}$ is the strength
tensor and $\widetilde{F}^{\mu\nu}=1/2\epsilon^{\mu\nu\rho\sigma}
F_{\rho\sigma}$ is its dual. This Chern-Simons term leads to the
rotations of the polarization vectors of photons when propagating
over the cosmological distance \cite{jackiw} \footnote{Such a term
also breaks the Einstein equivalence principle \cite{ni1,ni2} in the
short wavelength limit and breaks causality in the long wavelength
limit \cite{Lehnert:2004hq,Lehnert:2004be}.}. The change in the
position angle of the polarization plane $\Delta\chi$,
characterizing $CPT$ violation in this scenario, can be obtained by
observing polarized radiation from distant sources such as radio
galaxies,
 quasars \cite{jackiw,carroll}, and
the cosmic microwave background (CMB) \cite{kamionkowski,Feng:2004mq}. Assuming the rotation angle is
homogeneous and isotropic $\Delta\chi=\Delta\bar{\chi}$, the CMB power
spectra would be rotated as \cite{Feng:2004mq,cptv_boomrang}
 \bea\label{rotationformulas}
 C^{TT,obs}_l&=&C^{TT}_l~,\nonumber\\
  C^{TE,obs}_l&=&C^{TE}_l \cos{(2\Delta\bar{\chi})}~, \nonumber\\
 C^{TB,obs}_l&=&C^{TE}_l \sin{(2\Delta\bar{\chi})}~, \nonumber\\
 C^{EE,obs}_l&=&C^{EE}_l \cos^2{(2\Delta\bar{\chi})} +C^{BB}_l \sin^2{(2\Delta\bar{\chi})}~, \nonumber\\
   C^{BB,obs}_l&=&C^{EE}_l \sin^2{(2\Delta\bar{\chi})} +C^{BB}_l \cos^2{(2\Delta\bar{\chi})}~, \nonumber\\
  C^{EB,obs}_l&=&\frac{1}{2}\sin{(4\Delta\bar{\chi})} (C^{EE}_l-C^{BB}_l) ~.
  \eea
In the formulas above, the quantities with the superscript $obs$
are those observed after the rotation. $T$, $E$ and $B$ represent
the temperature, the electriclike and magneticlike polarization
modes, respectively.

In Ref. \cite{Feng:2004mq}, with Feng and Li, we did the
simulations on the measurement of $\Delta\bar{\chi}$ with the future
high precision CMB experiments, CMBPol \cite{cmbpol} and PLANCK
\cite{planck} using the rotation formulas (\ref{rotationformulas}).
We pointed out that in such experiments the $EB$ spectrum will be the
most sensitive probe of such $CPT$ violation, this is because the
$EB$ power spectrum is generated by the rotation of the $EE$ power
spectrum, which is a more sensitive probe of the primordial
fluctuations than the $TT$ and $TE$ spectra. In \cite{cptv_boomrang},
with Feng, Xia, and Chen, we first found that a nonzero
rotation angle $\Delta\bar{\chi}=-6.0\pm4.0$ deg ($1\sigma$) is
mildly favored by the CMB polarization data from the three-year
Wilkinson Microwave Anisotropy Probe (WMAP3) observations
\cite{Spergel:2006hy,Page:2006hz,Hinshaw:2006,Jarosik:2006,WMAP3IE}
and the January 2003 Antarctic flight of BOOMERanG (hereafter
B03)\cite{B03,B03EE,TCGC}  (see also
Ref.\cite{Liu:2006uh,Kostelecky:2007zz,Kostelecky:2008be,Geng:2007ga}).
This is a signal in some sense of the cosmological $CPT$ violation
mentioned above. Later on, Cabella, Natoli, and Silk \cite{cabella}  performed a
wavelet analysis of the temperature and polarization maps of the CMB
delivered by WMAP3. They set a limit on the rotation angle
$\Delta\bar{\chi}=-2.5\pm3.0$ deg ($1\sigma$).  This is consistent
with our result because they considered WMAP3 data only. Using the
full data of B03 and the WMAP3 angular power spectra, one of the
authors (X.Z.) with Xia $et$ $al$. \cite{xia1} has found that
$\Delta\bar{\chi}=-6.2\pm3.8$ deg ($1\sigma$). This result improved
the measurement given by our previous paper \cite{cptv_boomrang}.
Recently, the WMAP experiment has published the five-year results for
the CMB angular power spectra which include the $TB$ and $EB$
information \cite{WMAP51,WMAP52}. They used the polarization power
spectra of WMAP5, $TE/TB$ ($2\leq l \leq450$) and $EE/BB/EB$ ($2\leq
l \leq23$) to determine this rotation angle \cite{WMAPCPT} and
found that $\Delta\bar{\chi}=-1.7\pm2.1$ deg ($1\sigma$). However,
when B03 data are included, one of the authors (X.Z.) with Xia $et$ $al$.
\cite{xia2} found that $\Delta\bar{\chi}=-2.6\pm1.9$ deg
($1\sigma$). Again, a small $CPT$-violating effect is mildly
detected by current data.

We note that the rotation formulas given in (\ref{rotationformulas})
are valid only for a homogeneous and isotropic rotation angle and are obtained in
the Minkowski spacetime or the spatially flat Friedmann-Robertson-Walker
spacetime which is conformally equivalent to the former. This is expected to be 
a good approximation when the coupled scalar field $\phi$ is the dark energy or 
the function of the curvature scalar because in these cases $\phi$ is very homogeneous in the observed Universe, while the
accompanied perturbations are much smaller.  Usually, its background part makes the dominant contributions.
One of the aims of this paper is to study the secondary effects due to the perturbations of $\phi$, which leads to the anisotropies of the rotation angle. 
For this purpose, we first study 
the Maxwell theory modified by the Chern-Simons
term in the general curved spacetime and investigate the possibility of
obtaining the rotation angle in a fully general relativistic
framework. The spatial fluctuations of the scalar field
make the rotation angle inhomogeneous and anisotropic and bring higher order corrections to the rotation formulas Eq.
(\ref{rotationformulas}). Specifically,  we evaluate the magnitude of
these corrections in the models of tracking dark energy, as required
by the quintessential baryo/leptogenesis, and found the corrections
are negligible. However, for some other models, the corrections could
be sizable. This paper is organized as follows. In Sec. II, we
present the relevant equations of the modified electromagnetic
theory under the geometric optics approximation. In Sec. III, we
study the generalized Stokes parameters and the changes in CMB power
spectra in Sec. IV. In Sec. V, we evaluate the corrections due
to quintessence fluctuations in the quintessential baryogenesis
model. Section VI is the conclusion.

\section{Basic equations}

The full Lagrangian of the Maxwell theory modified by the Chern-Simons term (\ref{int}) (without other sources) is
\be\label{lagrangian}
\mathcal{L}=-{1\over 4}F_{\mu\nu}F^{\mu\nu}+
p_{\mu}A_{\nu}\widetilde{F}^{\mu\nu}~.
\ee
 This Lagrangian is not gauge-invariant, but
the action integral $S=\int \mathcal{L} d^4x$ is gauge-independent
because $p_{\mu}$ is defined in (\ref{int}) as the derivative of the
scalar field. The equation of motion can be obtained through varying
this Lagrangian with respect to $A_{\nu}$: \be
\nabla_{\mu}F^{\mu\nu}=-2 p_{\mu}
\widetilde{F}^{\mu\nu}~.\label{maxwell} \ee The right-hand side of
the above equation is brought by the Chern-Simons term.  But the
identity is unchanged: \be
\nabla_{\mu}F_{\rho\sigma}+\nabla_{\rho}F_{\sigma\mu}+\nabla_{\sigma}F_{\mu\rho}=0~.\label{identity}
\ee We will study these equations in a gauge-independent way though
it is easier to do it by choosing the Lorenz gauge \cite{Li:2006ss}.
For this purpose we make a differentiation to Eq. 
(\ref{identity}) and get \bea\label{master} \Box
F_{\rho\sigma}+2\nabla_{\rho}(p_{\mu} \widetilde{F}^{\mu}_{~\sigma})
-2\nabla_{\sigma}(p_{\mu} \widetilde{F}^{\mu}_{~\rho})
-[F^{\alpha}_{~\rho}R_{\alpha\sigma}-F^{\alpha}_{~\sigma}R_{\alpha\rho}-F^{\mu\alpha}R_{\alpha\mu\rho\sigma}]
=0~, \eea where $R_{\alpha\sigma}$ and $R_{\alpha\mu\rho\sigma}$ are
Ricci and Riemann tensors, respectively.

Since we are studying the light which propagates in the cosmological
scales, the geometric optics approximation (GOA) applies very well.
With this approximation, the solution to the equation of motion is
supposed to be \be\label{goa} F^{\mu\nu}=(a^{\mu\nu}+\epsilon
b^{\mu\nu}+\epsilon^2 c^{\mu\nu}+...)e^{iS/\epsilon}~, \ee where we
made a complexification to the electromagnetic field, but $\epsilon$
is a small real parameter and $S$ is a real function. This ansatz
means that the phase of the wave varies much faster than the
amplitude. We define the wave vector as \be k_{\mu}\equiv
\nabla_{\mu}S~, \ee which represents the travel direction of the
photon.

Substituting the ansatz (\ref{goa}) into Eqs. (\ref{master})
and (\ref{identity}) and dropping out the terms containing Ricci and
Riemann tensors, we have \bea\label{master2} &
&\Box(a_{\rho\sigma}+\epsilon
b_{\rho\sigma}+...)+\frac{2i}{\epsilon}k^{\mu}
\nabla_{\mu}(a_{\rho\sigma}+\epsilon b_{\rho\sigma}+...)
+\frac{i}{\epsilon}(\nabla_{\mu}k^{\mu}) (a_{\rho\sigma}+\epsilon
b_{\rho\sigma}+...)-\frac{1}{\epsilon^2}k_{\mu}k^{\mu}
(a_{\rho\sigma}+\epsilon b_{\rho\sigma}+...)\nonumber\\
&=&-2[ (\nabla_{\rho}p^{\mu})(\widetilde{a}_{\mu\sigma}+\epsilon \widetilde{b}_{\mu\sigma}+...)
+p^{\mu}(\nabla_{\rho}\widetilde{a}_{\mu\sigma}+\epsilon \nabla_{\rho}\widetilde{b}_{\mu\sigma}+...)
+\frac{ik_{\rho}}{\epsilon}p^{\mu}(\widetilde{a}_{\mu\sigma}
+\epsilon \widetilde{b}_{\mu\sigma}+...)]+2[\rho\leftrightarrow \sigma]
\eea
and
\be\label{identity2}
[\nabla_{\mu}(a_{\rho\sigma}+\epsilon b_{\rho\sigma}+...)+\frac{i}{\epsilon}k_{\mu}(a_{\rho\sigma}+
\epsilon b_{\rho\sigma}+...)]+[\rho\sigma\mu]+[\sigma\mu\rho]=0~.
\ee

At the leading order of the GOA, Eq. (\ref{identity2}) gives \be
k_{\mu}a_{\rho\sigma}+k_{\rho}a_{\sigma\mu}+k_{\sigma}a_{\mu\rho}=0
\ee which implies that $a_{\rho\sigma}$ should have the following
antisymmetric form: \be\label{solution1}
a_{\rho\sigma}=k_{\rho}a_{\sigma}-k_{\sigma}a_{\rho}~. \ee Then we
collect the terms of Eq. (\ref{master2}) at the orders
of $1/\epsilon^2$ and $1/\epsilon$, respectively. At the order of
$1/\epsilon^2$, we have \be\label{solution2}
k_{\mu}k^{\mu}=0~. \ee The propagation equation of $k^{\mu}$ can be
obtained by differentiating the above equation again:
\be\label{solution21}
0=\nabla_{\nu}(k_{\mu}k^{\mu})=2\nabla^{\mu}S\nabla_{\nu}\nabla_{\mu}S=
2\nabla^{\mu}S\nabla_{\mu}\nabla_{\nu}S
=2k^{\mu}\nabla_{\mu}k_{\nu}~. \ee This is a geodesic equation. The
vector $k^{\mu}$ defines an affine parameter $\lambda$ which
measures the distance along the light ray: \be k^{\mu}\equiv
\frac{dx^{\mu}}{d\lambda}~. \ee We can see from (\ref{solution21})
that $k^{\mu}$ is parallelly transported along the light curve
$x^{\mu}(\lambda)$. In other words, photons travel along null
geodesics. These results are the same as those of the standard
Maxwell theory. The modification due to the Chern-Simons term
appears at the order of $1/\epsilon$: \be\label{solution3}
\mathcal{D}a^{\nu}+\frac{\theta}{2}a^{\nu}=
-p_{\mu}\epsilon^{\mu\nu\rho\sigma}k_{\rho}a_{\sigma}~, \ee where we
have considered Eq. (\ref{solution1}) and defined the operator
$ \mathcal{D}\equiv k^{\mu}\nabla_{\mu}$. The quantity $\theta=
\nabla_{\mu}k^{\mu}$ describes the expansion of the bundle of the
light. Without the modification, the right-hand side of the above
equation would vanish. Its physical meaning is that the polarization
vector of the photon is not parallelly transported along the light
ray as we will see in the next section. In addition, by applying the GOA to
the original equation \be \nabla_{\mu}F^{\mu\nu}=-2
p_{\mu}\widetilde{F}^{\mu\nu} \ee we have \be\label{solution4}
k_{\mu}a^{\mu}=0~. \ee The basic results we got above are 
Eqs. (\ref{solution3}) and (\ref{solution21}) with two
orthogonality relations (\ref{solution2}) and (\ref{solution4}).

\section{Stokes parameters}

It is convenient to use the Stokes parameters to study the
polarization of radiation. The four Stokes parameters are well
defined in Minkowski spacetime (the inertial frame). Considering a
monochromatic electromagnetic wave of frequency $\omega_0$
propagating in the $+z$ direction \be
     E_x= a_x(t) \exp[i(\omega_0 t - \theta_x(t))],\qquad\qquad
     E_y = a_y(t) \exp[i(\omega_0 t - \theta_y(t))],
\ee
the Stokes parameters
are defined as the time averages
\bea\label{stokes}
     I & \equiv & \left\langle E_x E_x^{\ast}\right\rangle
                         +\left\langle E_y E_y^{\ast}\right\rangle ,\nonumber\\
     Q & \equiv & \left\langle E_x E_x^{\ast}\right\rangle
                         -\left\langle E_y E_y^{\ast}\right\rangle ,\nonumber\\
     U & \equiv & \left\langle E_x E_y^{\ast}\right\rangle
                         +\left\langle E_x^{\ast}E_y\right\rangle ,\nonumber\\
     V & \equiv & i[\left\langle E_x E_y^{\ast}\right\rangle
                         -\left\langle E_x^{\ast} E_y\right\rangle ].
\eea

In general relativity, these definitions should be generalized. This
can be done by using the tetrad formalism. A tetrad is a set of four
orthogonal unit basis vectors $e^{\mu}_{~(a)}$, with $a=0,1,2,3$. At
each point $x$, we can attach a tetrad which transforms between the
coordinate frame and the local inertial frame at $x$. For a vector
field $B_{\mu}(x)$, its components in the local inertial frame are
\be \bar{B}_{a}= e^{\mu}_{~(a)} B_{\mu}~. \ee The Latin indices are
lowered and raised by the Minkowski metric $\eta^{ab}$, the Greek
indices, however, by the coordinate metric $g^{\mu\nu}$. The tetrad
has the following properties: \be
g_{\mu\nu}e^{\mu}_{~(a)}e^{\nu}_{~(b)}=\eta_{ab}~,~~~\eta^{ab}e^{\mu}_{~(a)}e^{\nu}_{~(b)}=g^{\mu\nu}~.
\ee We can set the tetrad frame at each point as follows. Consider
the rest frame of the free fall observer, in which the four-velocity
is $\bar{u}^a=\delta^a_0$. Furthermore, we require the observer to
see the light traveling along the $+z$ direction, and hence
$\bar{k}^a=\omega (\delta^a_0+\delta^a_3)$. So, after transforming to the coordinate frame,  \be
u^{\mu}=e^{\mu}_{~(a)}\bar{u}^a=e^{\mu}_{~(0)}~, \ee and \be
k^{\mu}=e^{\mu}_{~(a)}\bar{k}^a=\omega (u^{\mu}+e^{\mu}_{~(3)})~.
\ee Hence, \be
e^{\mu}_{~(0)}=u^{\mu}~,~~~e^{\mu}_{~(3)}=\frac{1}{\omega}(k^{\mu}-\omega
u^{\mu})~, \ee where $\omega\equiv k_{\mu}u^{\mu}$ is the frequency
measured by the observer. The other tetrad vectors $e^{\mu}_{~(1)}$
and $e^{\mu}_{~(2)}$ are unit spacelike, orthogonal to each other
and to $e^{\mu}_{~(0)}$, $e^{\mu}_{~(3)}$, and therefore orthogonal to
$k^{\mu}$.

The electric vector in general spacetime for an observer with
four-velocity $u^{\mu}$ is defined as \be E^{\mu}\equiv
F^{\mu\nu}u_{\nu}~. \ee At the leading order of the GOA mentioned at the
last section, it is \be E^{\mu}=
a^{\mu\nu}u_{\nu}e^{iS/\epsilon}=(k^{\mu}a^{\nu}-k^{\nu}a^{\mu})u_{\nu}e^{iS/\epsilon}~.\ee
Transforming it to the local inertial frame, we get the $x$ and $y$
components of the electric field in this frame easily: \be
E_x=\bar{E}_1=E_{\mu} e^{\mu}_{~(1)}~,~~~E_y=\bar{E}_2=E_{\mu}
e^{\mu}_{~(2)}~. \ee In the local inertial frame, the definitions of
the Stokes parameters (\ref{stokes}) are applicable. By applying the
above equations to (\ref{stokes}), we get the general expressions of
the Stokes parameters in curved spacetime
\cite{anile,Kopeikin:2005jm}: \bea
I&=&\omega^2 L_{\mu\nu}(e^{\mu}_{~(1)} e^{\nu}_{~(1)}+e^{\mu}_{~(2)} e^{\nu}_{~(2)})\nonumber\\
Q&=&\omega^2 L_{\mu\nu}(e^{\mu}_{~(1)} e^{\nu}_{~(1)}-e^{\mu}_{~(2)} e^{\nu}_{~(2)})\nonumber\\
U&=&\omega^2 L_{\mu\nu}(e^{\mu}_{~(1)} e^{\nu}_{~(2)}+e^{\mu}_{~(2)} e^{\nu}_{~(1)})\nonumber\\
V&=&i\omega^2 L_{\mu\nu}(e^{\mu}_{~(1)}
e^{\nu}_{~(2)}-e^{\mu}_{~(2)} e^{\nu}_{~(1)})~, \eea where
$L_{\mu\nu}\equiv <a_{\mu}a_{\nu}^{\ast}>$ satisfies the following
equation making use of Eq. (\ref{solution3}): \be \mathcal{D}
L_{\mu\nu}+\theta
L_{\mu\nu}=-p_{\alpha}k_{\beta}(\epsilon^{\alpha\beta\gamma}_{~~~~\mu}
L_{\gamma\nu}+\epsilon^{\alpha\beta\gamma}_{~~~~\nu}L_{\mu\gamma})~.
\ee We can see that the Stokes parameters are coordinate scalars but
not Lorentz scalars. We require the tetrad frames to be not physically
rotating. In order to do that, we set the tetrad vectors at each
point so that $e^{\mu}_{~(1)}$ and $e^{\mu}_{~(2)}$ are parallelly
transported along the light curve. So it is straightforward to get
the propagation equations of the four parameters along the light
curve: \bea
& & \mathcal{D} F_0+\theta F_0=0~,\\
& & \mathcal{D} F_1+\theta F_1=2p_{\mu}k^{\mu} F_2~,\label{F1}\\
& & \mathcal{D} F_2+\theta F_2=-2p_{\mu}k^{\mu} F_1~,\label{F2}\\
& & \mathcal{D} F_3+\theta F_3=0~, \eea where $F_a\equiv
(I,~Q,~U,~V)/\omega^2$. Equation (33) means the conservation of
the light flux. Equation (36) indicates that the Stokes $V$,
which describes the net circular polarization, vanishes if it is
zero at the beginning. This is the case for CMB where the
polarization is produced at the last scattering. Since the Stokes
$V$ cannot be produced by Thomson scattering, it remains
zero afterwards. In short, the net circular polarization remains
vanishing even in the presence of the Chern-Simons term. The terms
in the right-hand sides of Eqs. (34) and (35) are the effects of the
Chern-Simons term which rotates the polarization angle of the light.
The polarization angle defined by $\chi\equiv 1/2 \arctan{(U/Q)}=1/2
\arctan{(F_2/F_1)}$ satisfies \be \mathcal{D}\chi+p_{\mu}k^{\mu}=0~.
\ee This angle when measured at the point $f$ is rotated by \be
\Delta\chi=\chi_{f}-\chi_{i}=-\int^{f}_{i}p_{\mu}k^{\mu}d\lambda=-\int^{f}_{i}p_{\mu}dx^{\mu}(\lambda)~,
\ee compared with that at the point $i$ when the photon was emitted.
From (\ref{int}),  $p_{\mu}=-(2c\alpha_{em})/(3\pi
M)\partial_{\mu}\phi$, the rotation angle is given by
 \be\label{angle}
\Delta\chi=\frac{2c\alpha_{em}}{3\pi M}(\phi_f-\phi_i)~. \ee
Defining \be F_{\pm}\equiv F_1\pm iF_2~, \ee it satisfies from
Eqs. (\ref{F1}) and (\ref{F2}) that \be F_{\pm}^f=F_{\pm}^i
\exp{(-\int^f_i\theta d\lambda)}\exp{(\pm i2\Delta\chi)}~. \ee The
Chern-Simons term modifies the result by merely adding the rotation
factor $\exp{(\pm i2\Delta\chi)}$. Hence the observed Stokes
parameters should be \be\label{ro} (Q\pm iU)^{obs}=\exp{(\pm
i2\Delta\chi)}(Q\pm iU)~. \ee This is the basic result obtained in
this section. It describes the rotation of the polarization of a
single bundle of light. It is the starting point to study the
rotated CMB power spectra in the next section.

\section{CMB Power Spectra}

In order to analyze the CMB map, we usually make multipole
expansion. In the flat Universe \footnote{For the treatment of CMB
anisotropies in open and close Universe, please see
\cite{Hu:1997mn}}, we can expand the temperature and polarization
anisotropies in terms of appropriate spin-weighted harmonic
functions on the sky \cite{Zaldarriaga:1996xe}: \bea
T(\hat{\bi{n}})&=& \sum_{lm}a_{T,lm}Y_{lm}(\hat{\bi{n}})\nonumber \\
(Q\pm iU) (\hat{\bi{n}})&=& \sum_{lm} a_{\pm 2, lm} \;_{\pm 2}Y_{lm}(\hat{\bi{n}})~.
\eea
The expressions for the expansion coefficients are
\begin{eqnarray}
a_{T,lm}&=&\int d\Omega\; Y_{lm}^{*}(\hat{\bi{n}}) T(\hat{\bi{n}})
\nonumber  \\
a_{\pm 2,lm}&=&\int d\Omega \;_{\pm 2}Y_{lm}^{*}(\hat{\bi{n}}) (Q\pm iU)(\hat{\bi{n}})~.\label{alm}
\end{eqnarray}
Instead of $a_{2,lm}$ and $a_{-2,lm}$, it is convenient to introduce their
linear combinations
\begin{eqnarray}
a_{E,lm}=-(a_{2,lm}+a_{-2,lm})/2 \nonumber \\
a_{B,lm}=i(a_{2,lm}-a_{-2,lm})/2.
\label{aeb}
\end{eqnarray}
The power spectra are defined as \be \langle a_{X',l^\prime
m^\prime}^{*} a_{X,lm}\rangle= C^{X'X}_{l} \delta_{l^\prime l}
\delta_{m^\prime m} \ee with the assumption of statistical isotropy.
In the equation above, $X'$ and $X$ denote the temperature $T$ and
the $E$ and $B$ modes of the polarization field, respectively. For Gaussian
theories, the statistical properties of the CMB
temperature/polarization map are specified fully by these six
spectra. In the standard case, $C^{TB}_l=C^{EB}_l=0$.

Considering the rotation in Eq. (\ref{ro}), the expressions for the
expansion coefficients become \be a_{\pm 2,lm}^{obs}=\int d\Omega
\:_{\pm 2}Y_{lm}^{\ast}(\hat{\bi{n}})(Q\pm iU)^{obs}(\hat{\bi{n}})
=\int d\Omega \:_{\pm 2}Y_{lm}^{\ast}(\hat{\bi{n}})\exp{(\pm
i2\Delta\chi)}(Q\pm iU)(\hat{\bi{n}})~, \ee and $a_{T,lm}$ remains
unchanged. The rotation angle in (\ref{angle}) depends on time as
well as space generally. It can be separated as the background part,
which is homogeneous  and isotropic, and the perturbation, which is randomly
distributed on the sky: \be
\Delta\chi=\Delta\bar{\chi}+\Delta\delta\chi~, \ee where \bea
& &\Delta\bar{\chi}=\frac{2c\alpha_{em}}{3\pi M}[\bar{\phi}(\eta_0)-\bar{\phi}(\eta_{dec})]\label{chi0}\\
& &\Delta\delta\chi=-\frac{2c\alpha_{em}}{3\pi
M}\delta\phi(\bi{x}_{dec},~\eta_{dec})~.\label{chi1} \eea In the above
equations the subscript $0$ indicates the present values and $dec$
means the values at the time of matter-radiation decoupling. The
homogeneous part $\Delta\bar{\chi}$ is the same one that appeared in the
previous rotation formulas (\ref{rotationformulas}). The final value
of the fluctuation $\delta\phi(\bi{x}_0,~\eta_0)$ is neglected
because it only gives rise to a dipole contribution due to our motion
with respect to the CMB frame.  In the flat Universe,
$\bi{x}_{dec}=(\eta_0-\eta_{dec})\hat{\bi{n}}$ when putting the
observer at the origin of the coordinate system. Similar to the
studies on Faraday rotation of the CMB polarization by a stochastic
magnetic field \cite{Kosowsky:2004zh}, we expand $\Delta\delta\chi$
on the sky: \be\label{chi2}
\Delta\delta\chi=\sum_{lm}b_{lm}Y_{lm}(\hat{\bi{n}})~, \ee and
define its angular power spectrum as \be\label{chi3} \langle
b_{l^\prime m^\prime}^{*} b_{lm}\rangle= C^{\chi}_{l}
\delta_{l^\prime l} \delta_{m^\prime m} ~, \ee where we have also
assumed statistical isotropy of $b_{lm}$. This angular power
spectrum is related to the power spectrum of $\delta\phi$ at time
$\eta_{dec}$, which can be seen from the following discussions.
Expanding $\delta\phi(\bi{x}_{dec},~\eta_{dec})$ in terms of Fourier
functions, we have \bea \delta\phi(\bi{x}_{dec},~\eta_{dec})&=&\int
\frac{d^3 k}{(2\pi)^{3/2}}\phi_{\bi{k}}(\eta_{dec})
e^{i\bi{k}\cdot\hat{\bi{n}} \Delta\eta}\nonumber\\
&=&\int \frac{d^3 k}{(2\pi)^{3/2}}\phi_{\bi{k}}(\eta_{dec})\sum_l(2l+1)i^lj_l(k\Delta\eta)
P_l(\hat{\bi{k}}\cdot\hat{\bi{n}})\nonumber\\
&=&\int \frac{d^3
k}{(2\pi)^{3/2}}\phi_{\bi{k}}(\eta_{dec})\sum_{lm}4\pi i^l
j_l(k\Delta\eta) Y^{\ast}_{lm}(\hat{\bi{k}})Y_{lm}(\hat{\bi{n}})~,
\eea where $\Delta\eta\equiv \eta_0-\eta_{dec}$, $j_l$ is the
spherical Bessel function and $P_l$ is the Legendre polynomial.
Comparing it with Eqs. (\ref{chi1}) and (\ref{chi2}), we get
\be b_{lm}=-\frac{8c\alpha_{em}}{3 M} i^l \int \frac{d^3
k}{(2\pi)^{3/2}}\phi_{\bi{k}}(\eta_{dec})j_l(k\Delta\eta)
Y^{\ast}_{lm}(\hat{\bi{k}})~. \ee With the help of the definition of
the power spectrum of $\delta\phi$, \be \langle
\phi^{\ast}_{\bi{k}'}(\eta_{dec})\phi_{\bi{k}}(\eta_{dec})\rangle\equiv
\frac{2\pi^2}{k^3}
\mathcal{P}_{\phi}(k,~\eta_{dec})\delta^3(\bi{k}-\bi{k}')~, \ee we
can find that $C^{\chi}_l$ in Eq. (\ref{chi3}) is
\be\label{chi4} C^{\chi}_l=\frac{16c^2\alpha^2_{em}}{9\pi M^2}\int
\frac{dk}{k} \mathcal{P}_{\phi}(k,~\eta_{dec}) j^2_l(k\Delta\eta)~,
\ee and \be \sum_{l}(2l+1)C^{\chi}_{l}=4\pi
\langle\Delta\delta\chi^2\rangle=\frac{16c^2\alpha^2_{em}}{9\pi
M^2}\langle \delta\phi^2\rangle~, \ee where
$\delta\phi=\delta\phi(\bi{x}_{dec},~\eta_{dec})$ and the arguments
$\bi{x}_{dec},~\eta_{dec}$ are suppressed in the following.

With these formulas, we can  calculate the coefficients after the rotation:
\bea
a_{\pm 2, lm}^{obs}&=&\int d\Omega \;_{\pm 2}Y_{lm}^{\ast}(\hat{\bi{n}}) (Q\pm iU)^{obs}(\hat{\bi{n}})\nonumber\\
&=&\exp{(\pm i2\Delta\bar{\chi})}\sum_{l_1m_1}a_{\pm 2, l_1m_1}\int d\Omega \;_{\pm 2}Y_{lm}^{\ast}(\hat{\bi{n}})
\exp{(\pm 2i \Delta\delta\chi)}\;_{\pm 2}Y_{l_1m_1}(\hat{\bi{n}})\nonumber\\
&=& \exp{(\pm i2\Delta\bar{\chi})}\sum_{l_1m_1}a_{\pm 2, l_1m_1}F^{\pm}_{lml_1m_1}~.
\eea
In the last equality, we have defined
\be
F^{\pm}_{lml_1m_1}\equiv \int d\Omega \;_{\pm 2}Y_{lm}^{*}(\hat{\bi{n}}) \exp{(\pm 2i \Delta\delta\chi)}\;_{\pm 2}
Y_{l_1m_1}(\hat{\bi{n}})~.
\ee
So
\bea
a^{obs}_{E,lm}&=&\frac{1}{2}\sum_{l_1m_1}[ (e^{i2\Delta\bar{\chi}}F^{+}_{lml_1m_1}+e^{-i2\Delta\bar{\chi}}
F^{-}_{lml_1m_1})a_{E,l_1m_1}+
i(e^{ i2\Delta\bar{\chi}}F^{+}_{lml_1m_1}-e^{-i2\Delta\bar{\chi}}F^{-}_{lml_1m_1})a_{B,l_1m_1}]~,\nonumber\\
a^{obs}_{B,lm}&=&\frac{1}{2}\sum_{l_1m_1}[ (-i)(e^{i2\Delta\bar{\chi}}F^{+}_{lml_1m_1}-e^{-i2\Delta\bar{\chi}}
F^{-}_{lml_1m_1})a_{E,l_1m_1}+
(e^{ i2\Delta\bar{\chi}}F^{+}_{lml_1m_1}+e^{-i2\Delta\bar{\chi}}F^{-}_{lml_1m_1})a_{B,l_1m_1}]~.
\eea

To calculate the observed correlations of $T$, $E$ and $B$, we make the following assumptions: (i) the rotation
field $\chi$ or $\phi$ is uncorrelated with the primordial $T$, $E$, and $B$ modes;
(ii) the rotation angle is small everywhere. In addition,
we have $C^{TB}_l=C^{EB}_l=0$ for primordial modes. Hence, we need only to calculate the following six correlations:
$\langle F^{\pm}_{lml'm'}\rangle$,  $\sum_{l_1m_1}C^{XX'}_{l_1}\langle F^{+\ast}_{l'm'l_1m_1}F^{+}_{lml_1m_1}\rangle$,
$\sum_{l_1m_1}C^{XX'}_{l_1}\langle F^{-\ast}_{l'm'l_1m_1}F^{-}_{lml_1m_1}\rangle$,
$\sum_{l_1m_1}C^{XX'}_{l_1}\langle F^{-\ast}_{l'm'l_1m_1}F^{+}_{lml_1m_1}\rangle$ and
$\sum_{l_1m_1}C^{XX'}_{l_1}\langle F^{+\ast}_{l'm'l_1m_1}F^{-}_{lml_1m_1}\rangle$.
Up to the quadratic order of $\Delta\delta\chi$, we have
\be
\langle F^{\pm}_{lml'm'}\rangle\simeq \langle 1\pm 2i \Delta\delta\chi-2\Delta\delta\chi^2\rangle
\delta_{ll'}\delta_{mm'}=(1-2\langle\Delta\delta\chi^2\rangle) \delta_{ll'}\delta_{mm'}~,
\ee
and
\bea
&&\sum_{l_1m_1}C^{XX'}_{l_1}\langle F^{+\ast}_{l'm'l_1m_1}F^{+}_{lml_1m_1}\rangle\nonumber\\
& \simeq & 
\sum_{l_1m_1}C^{XX'}_{l_1}\int d\Omega' d\Omega[1-4\langle\Delta\delta\chi^2\rangle+4\langle\Delta\delta\chi(\hat{\bi{n}'})
\Delta\delta\chi(\hat{\bi{n}})\rangle]
\;_2Y_{l'm'}(\hat{\bi{n}'})\;_2Y_{l_1m_1}^{\ast}(\hat{\bi{n}'})\;_2Y_{lm}^{\ast}(\hat{\bi{n}})
\;_2Y_{l_1m_1}(\hat{\bi{n}})\nonumber\\
&=&(1-4\langle\Delta\delta\chi^2\rangle)C^{XX'}_l\delta_{ll'}\delta_{mm'}+4\sum_{l_1m_1l_2m_2}C^{XX'}_{l_1}C_{l_2}^{\chi}
 \int d\Omega' d\Omega\;_2Y_{l'm'}(\hat{\bi{n}'})\;_2Y_{l_1m_1}^{\ast}(\hat{\bi{n}'})Y_{l_2m_2}^{\ast}(\hat{\bi{n}'})
 \;_2Y_{lm}^{\ast}(\hat{\bi{n}})\;_2Y_{l_1m_1}(\hat{\bi{n}})Y_{l_2m_2}(\hat{\bi{n}})~.\nonumber\\
 \eea
The remaining integrals in the above equation may be expressed in terms of the Wigner-3$j$ symbol
through the general relation \cite{Hu:2000ee}:
\begin{eqnarray}
&&
\int d\Omega \;_{s}Y_{lm}^{\ast} \;_{s_1}Y_{l_1m_1}\;_{s_2}Y_{l_2m_2}  =(-1)^{m+s}
\sqrt{(2 l+1)(2 l_1 + 1)(2 l_2+1) \over 4\pi}
 \wjma {l}{l_1}{l_2}{s}{-s_1}{-s_2}
 \wjma{l}{l_1}{l_2}{-m}{m_1}{m_2}\,,
\label{eqn:threey}
\end{eqnarray}
So
\be
\sum_{l_1m_1}C^{XX'}_{l_1}\langle F^{+\ast}_{l'm'l_1m_1}F^{+}_{lml_1m_1}\rangle\simeq (1-4\langle
\Delta\delta\chi^2\rangle)C^{XX'}_l\delta_{ll'}\delta_{mm'}
+\sum_{l_1l_2}C^{XX'}_{l_1}C_{l_2}^{\chi}\frac{(2l_1+1)(2l_2+1)}{\pi} \wjma {l}{l_1}{l_2}{2}{-2}{0}^2\delta_{ll'}
\delta_{mm'}~,
\ee
where we have used the orthogonality relation of the 3$j$ symbol
 \be
 \sum_{m_1m_2}(2l+1) \wjma{l}{l_1}{l_2}{m}{m_1}{m_2}  \wjma{l'}{l_1}{l_2}{m'}{m_1}{m_2}= \delta_{ll'}\delta_{mm'}~.
 \ee
Similarly,
we can find that
\be
\sum_{l_1m_1}C^{XX'}_{l_1}\langle F^{-\ast}_{l'm'l_1m_1}F^{-}_{lml_1m_1}\rangle=\sum_{l_1m_1}C^{XX'}_{l_1}
\langle F^{+\ast}_{l'm'l_1m_1}F^{+}_{lml_1m_1}\rangle~,
\ee
and
\bea
& &\sum_{l_1m_1}C^{XX'}_{l_1}\langle F^{-\ast}_{l'm'l_1m_1}F^{+}_{lml_1m_1}\rangle=\sum_{l_1m_1}C^{XX'}_{l_1}
\langle F^{+\ast}_{l'm'l_1m_1}F^{-}_{lml_1m_1}\rangle\nonumber\\
& &\simeq (1-4\langle\Delta\delta\chi^2\rangle)C^{XX'}_l\delta_{ll'}\delta_{mm'}
+\sum_{l_1l_2}(-1)^{L+1}C^{XX'}_{l_1}C_{l_2}^{\chi}\frac{(2l_1+1)(2l_2+1)}{\pi} \wjma {l}{l_1}{l_2}{2}{-2}{0}^2
\delta_{ll'}\delta_{mm'}~,
\eea
where $L=l+l_1+l_2$ and we have used the permutation property of the 3$j$ symbol
\be
 \wjma{l}{l_1}{l_2}{-m}{-m_1}{-m_2}= (-1)^L\wjma{l}{l_1}{l_2}{m}{m_1}{m_2}~.
 \ee

 Consequently, we obtained the rotation formulas of the power spectra
 \bea\label{rotationformulas2}
 C^{TT,obs}_l &=& C^{TT}_l~,\nonumber\\
 C^{TE,obs}_l&=&C^{TE}_l \cos{(2\Delta\bar{\chi})} (1-2\langle\Delta\delta\chi^2\rangle)~,\nonumber\\
 C^{TB,obs}_l&=&C^{TE}_l \sin{(2\Delta\bar{\chi})} (1-2\langle\Delta\delta\chi^2\rangle)~,\nonumber\\
 C^{EE,obs}_l&=&[C^{EE}_l \cos^2{(2\Delta\bar{\chi})} +C^{BB}_l \sin^2{(2\Delta\bar{\chi})} ]
 (1-4\langle\Delta\delta\chi^2\rangle)\nonumber\\
 &+&\sum_{l_1l_2} \wjma{l}{l_1}{l_2}{2}{-2}{0}^2\frac{(2l_1+1)(2l_2+1)}{2\pi}C^{\chi}_{l_2}\{[1+(-1)^{L+1}
 \cos{(4\Delta\bar{\chi})}]C^{EE}_{l_1} +[1+(-1)^{L}\cos{(4\Delta\bar{\chi})}]C^{BB}_{l_1}\}~,\nonumber\\
  C^{BB,obs}_l&=&[C^{EE}_l \sin^2{(2\Delta\bar{\chi})} +C^{BB}_l \cos^2{(2\Delta\bar{\chi})} ]
  (1-4\langle\Delta\delta\chi^2\rangle)\nonumber\\
 &+&\sum_{l_1l_2} \wjma{l}{l_1}{l_2}{2}{-2}{0}^2\frac{(2l_1+1)(2l_2+1)}{2\pi}C^{\chi}_{l_2}
 \{[1+(-1)^{L}\cos{(4\Delta\bar{\chi})}]C^{EE}_{l_1} +[1+(-1)^{L+1}\cos{(4\Delta\bar{\chi})}]C^{BB}_{l_1}\}~,\nonumber\\
  C^{EB,obs}_l&=&\frac{1}{2}\sin{(4\Delta\bar{\chi})} (C^{EE}_l-C^{BB}_l) (1-4\langle\Delta\delta\chi^2\rangle)\nonumber\\
 &+&\sin{(4\Delta\bar{\chi})}\sum_{l_1l_2} \wjma{l}{l_1}{l_2}{2}{-2}{0}^2\frac{(2l_1+1)(2l_2+1)}{2\pi}
 C^{\chi}_{l_2}(-1)^{L+1}(C^{EE}_{l_1}-C^{BB}_{l_1})~.
  \eea
In comparisons with those in Eq. (4), Eq. (69) included the
corrections from spatial fluctuations.

From Eq. (69), we can see firstly that $C^{TB,obs}_l$ and
$C^{EB,obs}_l$ are proportional to $\sin{(\Delta\bar{\chi})} $,
which vanish when $\Delta\bar{\chi}=0$. This is understandable
because $CPT$ is violated only by the background field. Second, we
find that \bea
& &\sum_l(2l+1)(C^{EE,obs}_l+C^{BB,obs}_l)\nonumber\\
&=&\sum_l(2l+1)(C^{EE}_l +C^{BB}_l )(1-4\langle\Delta\delta\chi^2\rangle)
+\sum_{ll_1l_2}\wjma{l}{l_1}{l_2}{2}{-2}{0}^2\frac{(2l+1)(2l_1+1)(2l_2+1)}{\pi}C^{\chi}_{l_2}
(C^{EE}_{l_1} +C^{BB}_{l_1} )\nonumber\\
&=&\sum_l(2l+1)(C^{EE}_l +C^{BB}_l )(1-4\langle\Delta\delta\chi^2\rangle)+4\langle\Delta\delta\chi^2\rangle
\sum_{l_1}(2l_1+1)(C^{EE}_{l_1} +C^{BB}_{l_1} )\nonumber\\
&=&\sum_l(2l+1)(C^{EE}_l +C^{BB}_l )~,\label{conservation} \eea
where we have used another  orthogonality relation of the 3$j$
symbol
 \be
 \sum_{l}(2l+1) \wjma{l}{l_1}{l_2}{-m_1-m_2}{m_1}{m_2}^2 = 1~.
 \ee
The equality in Eq. (\ref{conservation}) is the direct
consequence of invariance of $Q^2+U^2$ under the rotation
(\ref{ro}).

\section{The evaluation on the magnitude of corrections in the tracking dark energy model}

Equations (\ref{rotationformulas2}) indicated that the most
important corrections appear at the order of
$\langle\Delta\delta\chi^2\rangle$. In this section, we consider a
model for quantitative estimation on the corrections. Specifically,
we take the quintessential baryo/leptogenesis model as we mentioned
in the introduction. For such a model, we have \be
\langle\Delta\delta\chi^2\rangle=\frac{4c^2\alpha_{em}^2}{9\pi^2M^2}\langle
\delta\phi^2\rangle \sim \frac{10^{-5}}{M^2}\langle
\delta\phi^2\rangle~. \ee As was pointed out in
\cite{quin_baryogenesis}, to generate enough baryon number
asymmetry, the quintessence field $\phi$ should have tracking
behavior, which happens, for example, in the Albrecht and Skordis model \cite{as}. In
the following, we will evaluate $\langle\Delta\delta\chi^2\rangle$
in this model. We consider the perturbed metric in the Newtonian
gauge: \be ds^2=(1+2\Phi)dt^2-a^2(1-2\Phi)dx^idx^i~, \ee where
$t=\int ad\eta$ is the cosmic time and $\Phi$ is the gravitational
potential. The linear perturbation equation of the quintessence is
\be \ddot{\delta\phi}+3H\dot
{\delta\phi}-\frac{\nabla^2}{a^2}\delta\phi+V''(\phi)\delta\phi=4\dot\phi\dot\Phi-2V'\Phi~.
\ee In the above equation, the dot denotes the derivative with respect to
$t$. The general solution to this equation is decomposed into two
parts: the adiabatic mode and the isocurvature one. For the model of
quintessence with a tracking solution, the isocurvature perturbation
decays away quickly \cite{quin_baryogenesis,abramo}. We need only
calculate the adiabatic perturbation, which satisfies the adiabatic
condition \be \frac{\delta p}{\dot p}=\frac{\delta \rho}{\dot\rho}~,
\ee i.e., \be \Phi=\frac{d}{dt}(\frac{\delta\phi}{\dot\phi})~. \ee
With the equation above we can find that at the time of
matter-radiation decoupling (in the matter-dominated epoch) the
adiabatic perturbation of quintessence on large scales is \be
\delta\phi=\frac{2\dot\phi}{3H}\Phi=\frac{2}{\sqrt{3}}\sqrt{\Omega_{\phi}}M_{pl}\Phi~,
\ee where we have considered the exact tracking behavior of
quintessence, $w_{\phi}=w_m=0$ and the well-known result
$\Phi=constant$ in the matter-dominated era. The parameter
$\Omega_{\phi}\leq 10^{-2}$ \cite{bean,doran} is the density of
quintessence at this time, and $M_{pl}=1/\sqrt{8\pi G}\sim 10^{18}$GeV
is the reduced Planck mass. So, with $\langle \Phi^2\rangle\sim
10^{-10}$, we have \be \langle\Delta\delta\chi^2\rangle\sim
10^{-7}\frac{M_{pl}^2}{M^2}\langle\Phi^2\rangle\sim
10^{-17}\frac{M_{pl}^2}{M^2}~. \ee If
$\langle\Delta\delta\chi^2\rangle\ll 1$, the corrections in Eqs.
(\ref{rotationformulas2}) can be neglected safely, which happens for
the cut-off scale $M\gg 10^{-8} M_{pl}\sim 10^{10}$ GeV. In the
quintessential baryo/leptogenesis model, $M\sim 10^8 T_D$
\cite{quin_baryogenesis} where $T_D$ is the decoupling temperature
of lepton number violating interaction and is around $10^{11}$ GeV
\cite{lihong}.

\section{Conclusion}

In this paper, we have studied the effects of the interaction with
the derivative coupling of the scalar field to photons given by the
 Chern-Simons term in the general curved spacetime. Under the
geometric optics approximation, we have obtained the general form of
the rotation angle in a gauge-invariant method. We have calculated
the corrections brought by the spatial fluctuations of the scalar
field to the rotation formulas. These corrections exist due to the
dynamics of the scalar field\footnote{During the writing of this
paper, the paper \cite{skordis} appeared in the eprint arXiv, which has
some similarities with the calculations of corrections from the
fluctuations of the scalar field in this paper.}; however, they have not
been considered in the literature on the CMB data analysis. We have
estimated the magnitude of the corrections in a model of scalar
field for the quintessential baryo/leptogenesis scenario and
fortunately found that the corrections are very small and can be neglected
safely in the fit to the CMB data. The same techniques can be applied to the case of gravitational leptogenesis
in which the coupled scalar is the function of the gravitational field. 
Similar techniques can be developed to other cases 
in which the Chern-Simons parameter has other origins. For example, in a more complicated case, 
where the parameter is not statistically isotropic or even has no power spectrum, 
the space components of $p_{\mu}$ in Eq. (\ref{int}) will bring the correlations between $a_{T,l'm'}$ and $a_{E,lm}$ of 
different $l'm'$ and $lm$ and so on. These complications are beyond the scope of this paper, and we leave them in the future work.

\section{Acknowledgement}
M.L. is grateful to Sergei Kopeikin for useful correspondences and to Taotao Qiu and Jun-Qing Xia for helpful discussions.
This work is supported in part by National
Science Foundation of China under Grants No. 10533010
and No. 10675136, and the 973 program No.2007CB815401,
and by the Chinese Academy of Sciences under Grant
No. KJCX3-SYW-N2.

{}

\end{document}